\documentclass[conference]{IEEEtran}
\IEEEoverridecommandlockouts
\usepackage{cite}
\usepackage{amsmath,amssymb,amsfonts}
\usepackage{algorithmic}
\usepackage{graphicx}
\usepackage{textcomp}
\usepackage{xcolor}
\usepackage{subcaption}
\usepackage{multirow}

\def\BibTeX{{\rm B\kern-.05em{\sc i\kern-.025em b}\kern-.08em
    T\kern-.1667em\lower.7ex\hbox{E}\kern-.125emX}}
\begin{document}

\IEEEoverridecommandlockouts
\IEEEpubid{\begin{minipage}[t]{\textwidth}\ \\[10pt]
        \centering\normalsize{979-8-3315-8170-1/26/\$31.00 \copyright 2026 IEEE}
        \hspace{\columnsep}\makebox[\columnwidth]{\parbox{\columnwidth}{\vspace{6pt}\scriptsize \copyright 2026 IEEE. Personal use of this material is permitted. Permission from IEEE must be obtained for all other uses, in any current or future media, including reprinting/republishing this material for advertising or promotional purposes, creating new collective works, for resale or redistribution to servers or lists, or reuse of any copyrighted component of this work in other works. DOI: 10.1109/ICHQP67927.2026.11756316}}
\end{minipage}} 

\title{Challenges in the Proper Metrological Verification of Smart Energy Meters 
\thanks{This research was funded in whole or in part by National Science Centre,
Poland – 2024/55/D/ST7/00441. For the purpose of Open Access, the author has applied a~CC BY public copyright licence to any Author Accepted Manuscript (AAM) version arising from this submission.}
}

\makeatletter
\newcommand{\linebreakand}{%
  \end{@IEEEauthorhalign}
  \hfill\mbox{}\par
  \mbox{}\hfill\begin{@IEEEauthorhalign}
}
\makeatother

\author{
\IEEEauthorblockN{Antonio Bracale}
\IEEEauthorblockA{\textit{Department of Engineering} \\
University of Naples Parthenope\\
Naples, Italy} 
\and
\IEEEauthorblockN{Jakub Janowicz, Piotr Kuwa{\l}ek, Grzegorz Wiczy{\' n}ski}
\IEEEauthorblockA{\textit{Institute of Electrical Engineering and Electronics} \\
Pozna{\' n} University of Technology\\
Pozna{\' n}, Poland \\
piotr.kuwalek@put.poznan.pl}
}

\maketitle

\begin{abstract}
The most common instruments currently used to measure active/reactive energy and power quality indicators are smart energy meters (EM). Unfortunately, the verification of such meters is currently performed under ideal conditions or with simple signal models, which do not recreate actual states occurring in the power grid and do not ensure the verification of the properties of their signal chains. This paper presents challenges in proper metrological verification of smart EM. It presents existing legal and normative requirements and scientific research directions regarding these meters. Although the meters tested comply with the normative and legal requirements, the results reveal numerous imperfections in the signal and measurement chains for the selected test signal. Based on the results of the research results, further directions have been determined in the field of smart EM.
\end{abstract}


\section{Introduction}

Energy meters (EM) are among the most widely used meters. Currently designed and implemented meters are analog-digital electronic devices with advanced measurement capabilities, classified as advanced metering infrastructure (AMI) systems or smart EM. Smart EM directly measure voltages and currents. These data are used to determine and record energy values and other quantities, such as active and reactive power, root-mean-square (rms) values of voltage and current, frequency, total harmonic distortion (THD), selected harmonics, and short-term ($P_{st}$) and long-term ($P_{lt}$) flicker indicators. Smart EM constitute a distributed source of information in the power grid, enabling, among other things, the acquisition of information on energy flows and power quality (PQ) at customer supply points. Real-time knowledge of the status of the power grid is particularly important in the presence of distributed energy sources. Smart EM are an indispensable diagnostic tool making this acquisition feasible. On the contrary, the use of PQA seems unattainable due to the large number of measurements and the need to conduct them at the supplier-customer connection. Furthermore, some customer power lines are designed so that it is impossible to take measurements at the supplier-customer connection. 

The widespread use of such meters and competition among manufacturers forces the drive for low prices. The complexity of smart EM (software and hardware) and the drive for low prices are somewhat contradictory. It is especially important to remember that any performance imperfections in smart EM can occur in a very large number of cases. Due to the dispersion of meters, detecting malfunctions is difficult. This requires exceptional care when designing and performing comprehensive smart EM tests. Laboratory experience with smart EM~\cite{b1,b2,b3,b3a,b4,b5,b6} and observations of their operation in real power grids demonstrate the need to expand testing beyond standard requirements. Smart EM with PQ evaluation functionality are the ``first-contact doctor" in the process of assessing the condition of the network. Any measurement and recording imperfections in smart meters, even minor ones, are replicated at the system-wide level. Therefore, it is important to thoroughly examine the properties of the smart meter, going beyond standard tests. Furthermore, their diagnostic capabilities must be improved and expanded. Taking all of the above into account, it is necessary to develop a test procedure for smart EM to verify energy measurement errors and indicators that determine the quality of power using smart EM, considering conditions that occur in real power networks.

\subsection{Contribution}

Unlike previous studies that mainly report meter behavior for specific loads or isolated power-quality phenomena, this paper proposes a metrologically-oriented verification perspective for smart EM. Its contribution lies in showing that a simple and analytically defined test signal can still reveal significant imperfections in the measurement and signal-processing chains of commercially available meters under conditions representative of modern grids. Therefore, the paper does not provide additional experimental evidence of meter inaccuracies, but proposes to identify a practical route toward defining a minimal set of reference test signals that can be both suitable for rigorous metrological assessment and effective in detecting reliability issues under realistic operating conditions.

To this aim, the paper presents laboratory selected test results for smart EM from various manufacturers, which demonstrate significant discrepancies in the measurements of signals representing states occurring in the actual power grid. It demonstrates that a simple test signal can identify numerous problems in smart EM measurements of selected quantities, likely resulting from oversimplification of measurement signal chains to ensure a relatively low price compared to dedicated, specialized measuring instruments. In addition, it evidences that a single test signal, which has been shown to be useful in verifying a flickermeter built into smart EM~\cite{b4}, can also be a reliable test signal for verifying other measurement quantities. The test results support the need to develop a new verification procedure for smart EM, whose implementation will increase the reliability of measurements performed by such meters.


\section{Current technical and scientific conditions regarding smart energy meters}\label{sII}

The basic IEC (International Electrotechnical Commission) standards for {AC} (Alternating Current) {EM} include the standard {IEC} 62052-11~\cite{b7}, which contains general requirements for EM and standards for the measurement of active energy (the standard IEC 62053-21~\cite{b8} for directly/indirectly connected classes 1 and 2 meters and the standard IEC~62053\mbox{-}22~\cite{b9} for indirectly connected classes 0.2S and 0.5S meters) and reactive energy (the standard IEC 62053-23~\cite{b10} for directly/indirectly connected classes 2 and 3 meters and the standard IEC 62053-24~\cite{b11} for directly/indirectly connected classes 0.5S, 1S and 1 meters) by such meters. Taking into account specific European requirements, CENELEC published an additional series of standards EN 50470. Building on the standards of IEC 62052/62053, the EN 50470 series of standards adds specific aspects required by the Measuring Instruments Directive (MID). These additions include: different accuracy classes and terminology, some specific accuracy requirements, and data and software protection requirements. For smart EM, the relevant standard is EN 50470-1~\cite{b12}, which contains general requirements for meters, and EN~50470\mbox{-}3~\cite{b13}, which covers active energy measurements for meters of class A, B, and C. In turn, for the measurement of parameters defining energy quality by smart EM, the basic standards for measuring these indicators apply, i.e. the standard IEC 61000-4-30~\cite{b14} for all indicators, where for the measurement of the degree of distortion, it refers to the standard IEC 61000-4-7~\cite{b15} and for the measurement of flicker severity caused by voltage fluctuations, it refers to the standard IEC 61000-4-15~\cite{b16}. For these indicators, European requirements are generally consistent with the provisions of IEC standards. In addition to the normative requirements, individual countries introduce additional legal regulations regarding smart EM. For example, in Poland, there is a regulation presented by the relevant ministry~\cite{b17}, which contains detailed guidelines for smart EM installed at individual points in the power network (for example, the annex to the indicated regulation presents a set of PQ parameters that must be measured by EM in addition to active and reactive energy). 

The current technical specifications for smart EM presented herein contain simplified signal models that do not encompass the states (disturbances) that occur in a real power grid. As a result, verifying the compliance of smart EM with the standards cited does not ensure reliable measurements in real-world conditions. For example, the standard IEC~62053\mbox{-}24~\cite{b11} specifies tests in which test signals contain only a 5-$th$ order harmonic with a specific amplitude. In reality, voltage and current signals contain a set of harmonics that define the quasi-steady state shape. Furthermore, it is worth noting that many disturbances in PQ occur simultaneously in modern power grids, such as ``clipped sine" voltage distortion~\cite{b18} as a result of the input stages of switched-mode power supplies, and voltage fluctuations~\cite{b19} resulting from sudden load changes in the power network. Therefore, the scientific literature contains numerous research results analyzing meter readings for conditions occurring in the power grid. For example, tests are carried out at points beyond the normative limits~\cite{b3,b5,b6} or for specific real-world loads supplied under controlled laboratory conditions~\cite{b20,b21,b22,b23}, or under experimental conditions in a real power grid~\cite{b24,b25,b26}. Analyzing readings for specific load operating conditions reveals the problem of meter reading reliability. However, it is not sufficient to determine a procedure for the proper metrological evaluation of smart EM. Therefore, it is necessary to develop a group of as few test signals as possible, taking into account the limited duration of equipment verification in appropriate and qualified laboratories, which represent conditions occurring in the real power grid or allow for the assessment of the measurement chain properties in relation to such conditions while maintaining the possibility of metrological assessment.

\section{Selected research results and discussion}\label{sIII}

Section~\ref{sII} indicates that EM testing should be performed using signals that recreate real states (including harmonics, ``clipped sine" distortions and voltage fluctuations), or using signals that are as simple as possible, enabling metrological assessment (thanks to the simple and explicit form of the equation describing the test signal) and enabling verification of the largest possible number of measurement signal chains. 

This section presents selected results for an example signal that, on the one hand, has a simple functional description, which facilitates metrological evaluation and, on the other hand, allows for the properties of evaluation of measurement chain in relation to potential conditions that can occur in a power grid. The test signal can be expressed as follows:
\begin{equation}
u(t) = \sqrt{2} U_c \cos{(2\pi f_c t) + u_i^* \cos{(2\pi f_i t)}}\label{eq}
\end{equation}
\noindent
where $U_c$ is the rms value of the fundamental component at the fundamental frequency $f_c$, and $u_i^*$ is the relative value of the component at the frequency $f_i$ (subharmonic/interharmonic/harmonic component). It is worth noting that, in reality, signals found in modern power grids can be more complex. However, identifying the problem for an idealized two-component signal indicates a potentially even greater problem for a more complex signal. However, the simplicity of the signal used allows its widespread use in calibration and accreditation laboratories. 

\begin{figure}[htbp]
\centerline{\includegraphics[width=.8\columnwidth]{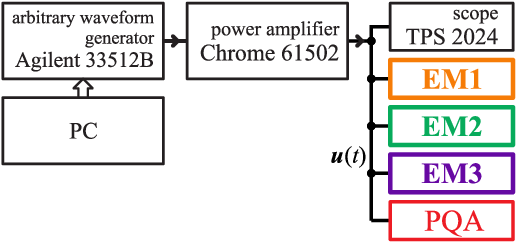}}
\caption{Diagram of the laboratory setup.}
\label{fig1}
\end{figure}

Using the indicated signal, tests were conducted according to the diagram shown in Fig.~\ref{fig1}. The generated signal was amplified to a level where $U_c$ was equal to 230~V, that is, the nominal rms value in the low-voltage network. The generated and amplified test signal was then fed to the voltage inputs of a PQ BOX 100 class A power quality analyzer (PQA) and smart EM from three different manufacturers (EM1/EM2/EM3). PQ parameters selected for the study are $P_{st}$, THD and $f_c$. For the smart EM tested, no deviations have been observed for the rms value and negative sequence content measurements; therefore, due to page limits, these results are not presented. The tested smart EM do not record other PQ parameters. The class A PQA is chosen as the reference control meter because it measures the indicators presented in this paper in accordance with the stringent metrological requirements specified in the standard IEC 61000-4-30~\cite{b14}.

\begin{figure}[htbp]
\centerline{\includegraphics[width=.7\columnwidth]{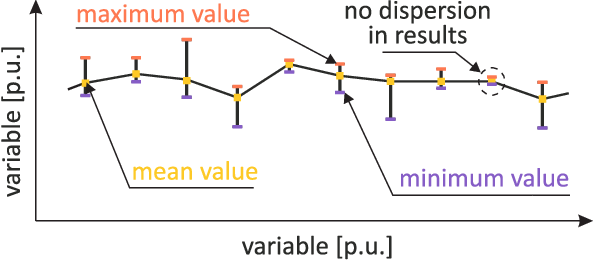}}
\caption{The method of presenting measurement results.}
\label{fig1a}
\end{figure}

\noindent
For each test signal setting, nine measurement results were obtained for a single meter manufacturer, because the tests used three separate meters from each manufacturer, and the results were obtained for each phase separately (three-phase meters were used, with the same test signal applied to each phase). On the basis of the results obtained from the meters, statistics were determined in the form of maximum, minimum, and mean values. The presentation of statistically processed measurement results of smart EM is shown in Fig.~\ref{fig1a}, where mean values are connected by a line and marked with a square marker.

\begin{figure}[htbp]
\centerline{\includegraphics[width=.9\columnwidth]{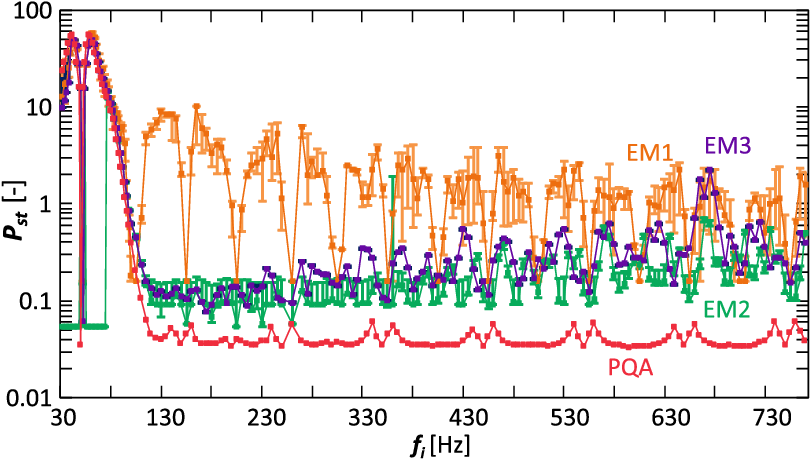}}
\caption{Characteristics of the $P_{st}$ values determined by individual smart EM (EM1/EM2/EM3) and by the PQA as a function of $f_i$.}
\label{fig2}
\end{figure}

Fig.~\ref{fig2} presents the characteristics of the minimum and maximum $P_{st}$ values determined by individual smart EM (EM1/EM2/EM3) and by the PQA as a function of the frequency $f_i$ of the additional component of the test signal. On the obtained results, significant discrepancies can be observed between the measurement results of the advanced measurement infrastructure, such as the class A PQA, and the results obtained for individual smart EM. It should be noted that the results obtained by the class A PQA are consistent with those obtained by analytical calculations, as shown in~\cite{b4}. For various $f_i$ ranges, the discrepancies indicate that the limit values in the power network~\cite{b27} exceeded, although no flicker severity caused by voltage fluctuations occurred. These discrepancies are probably the result of an inappropriate sampling frequency and the failure to take into account the transition bandwidth of the antialiasing filter used~\cite{b4,b5}.

\begin{figure}[hbtp]
\centering
\begin{subfigure}[b]{.9\columnwidth}
\centering
\includegraphics[width=\columnwidth]{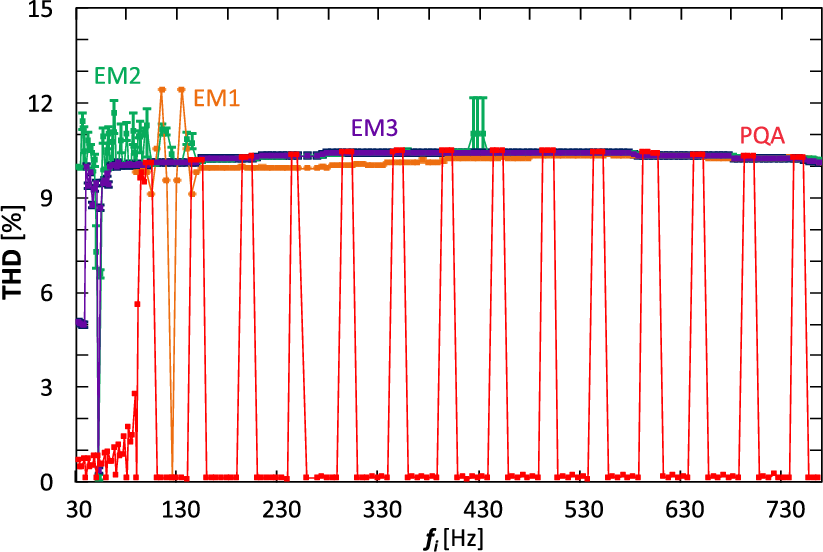}
\caption{all range}
\end{subfigure}
\begin{subfigure}[b]{.9\columnwidth}
\centering
\includegraphics[width=\columnwidth]{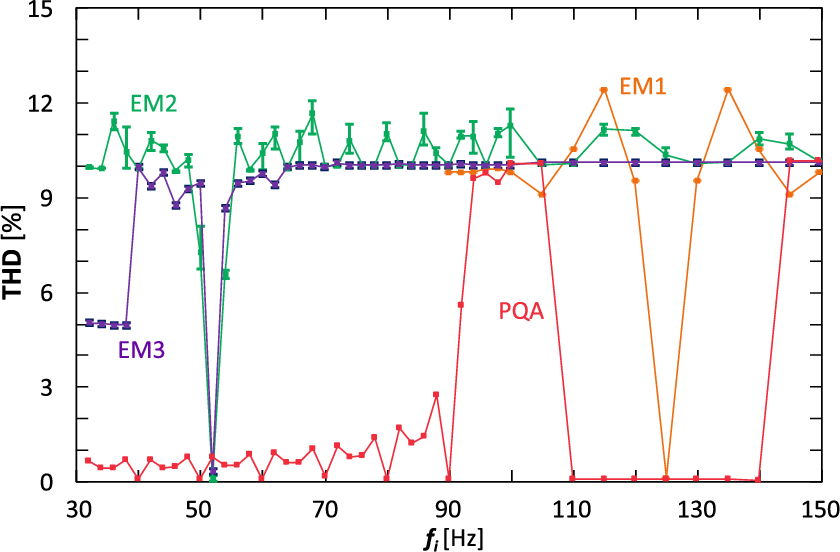}
\caption{zoom in the range of 30 to 150~Hz}
\label{fig3:b}
\end{subfigure}
\caption{Characteristics of THD values determined by individual smart EM (EM1/EM2/EM3) and by PQA as a function of $f_i$.}
\label{fig3}
\end{figure}

Fig.~\ref{fig3} present the characteristics of the minimum and maximum THD values determined by individual smart EM (EM1/EM2/EM3) and by the PQA as a function of the frequency $f_i$ of the additional component of the test signal. For frequencies $f_i$ close to harmonic frequencies (signal distortion caused by higher harmonics), one can observe the consistency of the PQA results and the individual EM. For the remaining frequency $f_i$ values, significant discrepancies can be observed between the readings of the advanced measurement infrastructure, such as the class A PQA, and the results obtained for the individual smart EM. It is worth noting that the results obtained by the class A PQA are consistent with those obtained by analytical calculations~\cite{b3,b3a}. These discrepancies are probably related to different implementations of the THD determination algorithm on smart EM processors~\cite{b3,b3a}, and the applied fundamental component estimation method, where the results may be influenced by the properties of the filter or frequency analysis algorithm used. It is worth paying attention to the significant variation of THD measurement results for individual smart EM, noticeable in the bandwidth for $f_i$ up to 150~Hz, as shown in Fig.~\ref{fig3:b}. The variation in this range is probably due to the problem of estimating the fundamental component for THD calculations, caused by the presence of an additional component with a frequency close to the fundamental frequency, which affects the THD determination.

Fig.~\ref{fig4} presents the characteristics of the minimum and maximum $f_c$ values determined by individual smart EM (EM1/EM2/EM3) and by the PQA as a function of the frequency $f_i$ of the additional component of the test signal. The results for smart EM (EM1/EM2) are comparable to the results for the PQA. In the case of one manufacturer’s model (EM3), significant discrepancies can be observed, exceeding permissible limits in 99.5\% of the week in the power grid~\cite{b27}, even though the fundamental frequency~$f_c$ was constant at 50~Hz (the nominal frequency value in the European power grid). The observed discrepancies may result from the fact that this indicator in this model is determined using the zero-loop method with a fundamental component filter applied. Poor filter quality may be the cause of these discrepancies.

\begin{figure}[htbp]
\centerline{\includegraphics[width=.9\columnwidth]{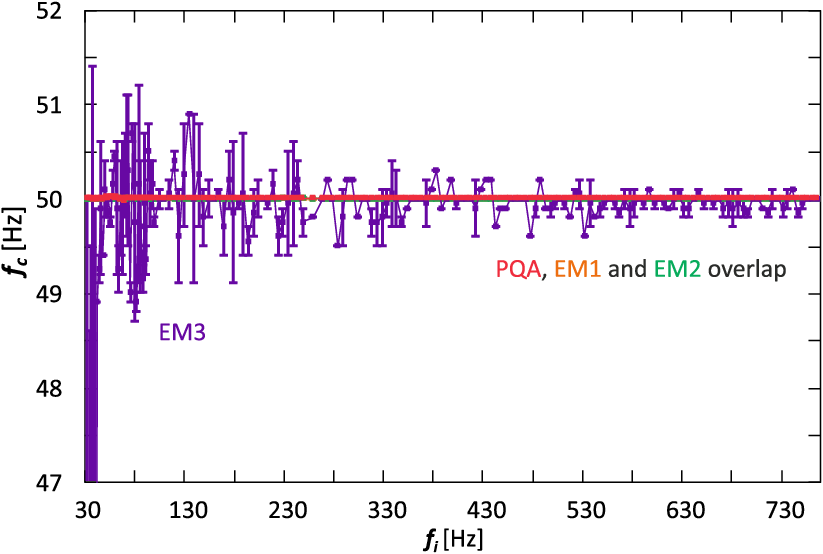}}
\caption{Characteristics of the $f_c$ values determined by the individual smart EM (EM1/EM2/EM3) and by the PQA as a function of $f_i$.}
\label{fig4}
\end{figure}

To better illustrate the differences between the individual instruments analyzed, the magnitude of the absolute error ~$\Delta \textnormal{AI}$ and the range of the individual measurement results $R$ were determined. The value of~$\Delta \textnormal{AI}$ is given by the equation:
\begin{equation}
\Delta \textnormal{AI} = |\textnormal{AI}_{\textnormal{EM}}-\textnormal{AI}_{\textnormal{PQA}}|,
\label{eq2}
\end{equation}
\noindent
where $\textnormal{AI}_{\textnormal{EM}}$ is the specific analyzed indicator measured by the smart EM, and $\textnormal{AI}_{\textnormal{PQA}}$ is the reference value of the analyzed indicator measured by a class A PQA. The range of the individual measurement results~$R$ is given by the equation:
\begin{equation}
R=\textnormal{AI}_{\textnormal{EM}}^{\textnormal{max}}-\textnormal{AI}_{\textnormal{EM}}^{\textnormal{min}},
\label{eq3}
\end{equation}
where $\textnormal{AI}_{\textnormal{EM}}^{\textnormal{max}}$ and $\textnormal{AI}_{\textnormal{EM}}^{\textnormal{min}}$ are the maximum and minimum values measured by the smart EM, respectively. Table~\ref{tab1} presents the maximum values of $\Delta \textnormal{AI}$ and $R$ in the individual $f_i$ ranges. Analyzing the results summarized in Table~\ref{tab1}, it can be seen that the largest discrepancies with the class A PQA, as well as the greatest variations in the readings of individual smart EM, occur for the $f_i$ range up to 100~Hz. This discrepancy in readings may be related to the analog filter used for the fundamental component, whose transition bandwidth may vary depending on the manufacturing dispersion of the filter components.

\begin{table}[htb]
\centering
\caption{Summary of the model of the maximum values of $\Delta \textnormal{AI}$ and $R$ in the individual $f_i$ ranges}
\label{tab1}
\begin{tabular}{c|c|ccc|ccc}
\multirow{2}{*}{AI}           & \multirow{2}{*}{$f_i$ {[}Hz{]}} & \multicolumn{3}{c|}{$\Delta \textnormal{AI}$ {[}p.u.{]}} & \multicolumn{3}{c}{$R$ {[}p.u.{]}} \\ \cline{3-8} 
                              &                              & EM1        & EM2        & EM3       & EM1       & EM2      & EM3       \\ \hline
\multirow{8}{*}{\rotatebox[origin=c]{90}{$P_{st}$ {[}-{]}}}  & {[}0;100)                    & 15.27      & 55.17      & 35.92     & 17.14     & 0.30     & 35.64     \\
                              & {[}100;200)                  & 10.30      & 0.13       & 0.10      & 6.25      & 0.10     & 0.01      \\
                              & {[}200;300)                  & 6.91       & 0.13       & 0.20      & 4.98      & 0.10     & 0.02      \\
                              & {[}300;400)                  & 4.06       & 1.86       & 0.32      & 3.46      & 1.81     & 0.00      \\
                              & {[}400;500)                  & 3.80       & 0.35       & 0.53      & 3.21      & 0.08     & 0.00      \\
                              & {[}500;600)                  & 2.72       & 0.56       & 0.59      & 2.04      & 0.09     & 0.01      \\
                              & {[}600;700)                  & 2.64       & 0.68       & 2.25      & 2.03      & 0.09     & 0.01      \\
                              & {[}700;800)                  & 2.39       & 0.57       & 0.90      & 2.07      & 0.10     & 0.01      \\ \hline
\multirow{8}{*}{\rotatebox[origin=c]{90}{THD {[}\%{]}}} & {[}0;100)                    & 11.95      & 11.42      & 11.25     & 0.10      & 1.53     & 10.00     \\
                              & {[}100;200)                  & 10.05      & 11.26      & 10.15     & 0.10      & 0.51     & 0.00      \\
                              & {[}200;300)                  & 10.16      & 10.31      & 10.35     & 0.10      & 0.01     & 0.10      \\
                              & {[}300;400)                  & 10.26      & 10.36      & 10.36     & 0.10      & 0.13     & 0.00      \\
                              & {[}400;500)                  & 10.35      & 12.10      & 10.36     & 0.10      & 1.74     & 0.00      \\
                              & {[}500;600)                  & 10.25      & 10.36      & 10.36     & 0.10      & 0.06     & 0.10      \\
                              & {[}600;700)                  & 10.15      & 10.25      & 10.25     & 0.10      & 0.08     & 0.10      \\
                              & {[}700;800)                  & 10.05      & 10.15      & 10.15     & 0.10      & 0.08     & 0.10      \\ \hline
\multirow{8}{*}{\rotatebox[origin=c]{90}{$f_c$ {[}Hz{]}}}  & {[}0;100)                    & 0.03       & 0.03       & 50.02     & 0.01      & 0.01     & 51.40     \\
                              & {[}100;200)                  & 0.01       & 0.01       & 0.90      & 0.01      & 0.01     & 1.80      \\
                              & {[}200;300)                  & 0.01       & 0.01       & 0.60      & 0.01      & 0.01     & 1.10      \\
                              & {[}300;400)                  & 0.00       & 0.00       & 0.50      & 0.00      & 0.00     & 0.70      \\
                              & {[}400;500)                  & 0.00       & 0.00       & 0.40      & 0.00      & 0.00     & 0.60      \\
                              & {[}500;600)                  & 0.00       & 0.00       & 0.40      & 0.00      & 0.00     & 0.50      \\
                              & {[}600;700)                  & 0.00       & 0.00       & 0.30      & 0.00      & 0.00     & 0.40      \\
                              & {[}700;800)                  & 0.00       & 0.00       & 0.20      & 0.00      & 0.00     & 0.30      \\ 
\end{tabular}
\end{table}

Analyzing sample results for $P_{st}$, THD, and $f_c$ as a function of the frequency $f_i$ of the additional component of the test signal, it can be seen that using a simple test signal allows for the identification of a number of irregularities in the operation of the tested smart EM. On the one hand, a simple test signal allows for the analytical calculation of reference indicators, and based on the known properties of the source used, it is possible to determine the uncertainty budget of the set value, which is particularly important from a metrological perspective. On the other hand, the readings of advanced instrumentation, such as a class A PQA with a specific inaccuracy, can be used as a reference value and compared with the readings obtained for smart EM. If the confidence intervals for the unknown measured quantities overlap, it can be assumed that the readings of the tested smart EM are correct. The results presented show that in practice there are discrepancies in the readings that significantly exceed the permissible limit values of selected parameters in the power network, whereas the analyzed test signal for the selected $f_i$ ranges did not cause the analyzed PQ disturbances (voltage fluctuations, voltage distortion, frequency fluctuations). 

\section{Conclusion}\label{sIV}

Currently, smart EM are among the most commonly used active/reactive EM and PQ indicators. Due to their widespread use, they are much cheaper than advanced measurement instruments, such as class A PQA. As a result, smart EM can exhibit significant imperfections in their measurement and signal chains. The paper points out that currently available smart EM are tested under ideal conditions or with simple test signals, which do not recreate the actual states occurring in modern power grids or allow for the effective identification of imperfections in their measurement and signal chains. It is worth noting that the growing share of renewable energy sources and electrification in modern power grids often leads to the simultaneous occurrence of multiple disturbances of PQ. Despite these conditions, measuring devices are expected to provide reliable measurements. 

This paper presents the results for an example test signal. Although it does not perfectly recreate the actual conditions that occur in modern power grids, it allows the identification of imperfections in the measurement and signal chains of smart EM, which may occur under real-world operating conditions with the simultaneous occurrence of multiple disturbances in PQ. Due to its simple functional description, the selected test signal can be used for metrological verification of the measurement results. The test results show significant discrepancies in smart EM readings and incompatibility with the current specialized equipment used. These discrepancies and variations are not specific to a single faulty smart EM from a single manufacturer but occur across multiple smart EM from different manufacturers. The presented research results demonstrate the necessity and challenges in finding the smallest possible number of test signals that will enable accurate metrological verification of smart EM and thus increase the reliability of these meters' measurement results in real-world conditions.


\vspace{12pt}

\end{document}